\documentclass[preprint,prc,showpacs,preprintnumbers,amsmath,amssymb,floatfix]{revtex4-2}
\usepackage{amsmath}
\usepackage{graphicx}
\usepackage{dcolumn}
\usepackage{bm}
\usepackage{ulem} 
\usepackage[usenames]{color}
\usepackage{epstopdf}
\usepackage{epsfig}
\usepackage{float}
\usepackage{subfigure}
\usepackage{multirow}
\usepackage{ulem}
\usepackage[colorlinks,linkcolor=blue,anchorcolor=teal,citecolor=red]{hyperref}

\begin{document}
	
	\preprint{}
	
	\title{Crossover Equation of State Constrained by Astronomical Observations and pQCD}

	\author{Xuesong Geng}
	
	\author{Kaixuan Huang}%
	\author{Hong Shen}%
    \thanks{Contact author:songtc@nankai.edu.cn}
	\author{Lei Li}%
	\thanks{Contact author:lilei@nankai.edu.cn}%
	\affiliation{%
		School of Physics, Nankai University, Tianjin 300071,  China
	}%
	
	\author{Jinniu Hu}
	\thanks{Contact author:hujinniu@nankai.edu.cn}%
	\affiliation{School of Physics, Nankai University, Tianjin 300071,  China}
	\affiliation{Shenzhen Research Institute of Nankai University, Shenzhen 518083, China}

	\date{\today}

	\date{\today}
	
	\begin{abstract}
			The hadron--quark crossover equation of state (EOS) of neutron star (NS) matter is investigated by combining relativistic mean-field (RMF) hadronic models with the Nambu--Jona-Lasinio (NJL) model for quark matter. The vector and diquark coupling constants of the NJL model are constrained using perturbative QCD (pQCD) calculations at high density through a scale-averaging likelihood approach, together with constraints from NS observations and the causality condition on the speed of sound. It is found that the diquark coupling is tightly constrained to $H \simeq 1.5G_s$, while the vector coupling is restricted to $G_v \lesssim 1.1G_s$ by the combined pQCD and astrophysical constraints. Crossover EOSs are constructed based on three hadronic RMF parameter sets, and their thermodynamic properties, sound speed behaviour, and trace anomaly are analysed. The resulting EOSs are applied to calculate NS global and dynamical properties, including mass--radius relations, tidal deformabilities, and fundamental radial oscillation frequencies. Compared with pure hadronic EOSs, the hadron--quark crossover is shown to significantly enhance the maximum NS mass, particularly for softer hadronic EOSs, while remaining consistent with observational bounds. It is further shown that the fundamental radial oscillation frequencies predicted by different EOSs exhibit pronounced differences, especially for intermediate-mass NSs, indicating that radial modes may provide a sensitive probe of the internal composition of NSs. These results indicate that quantitative NS observables may provide potential signatures of quark matter in NS interiors.
			
	\end{abstract}
	
	\maketitle
	
	
	\section{\label{sec:level1}INTRODUCTION}
   Neutron stars (NSs) are stellar remnants formed in the supernova explosions of massive stars. The central density of NSs can reach about five to six times the nuclear saturation density ($n_0 \sim 0.15~\mathrm{fm^{-3}}$) \cite{Lattimer2004}. Strong magnetic fields and rapid rotation lead to periodic electromagnetic signals that can be detected  by radio telescopes on Earth. These properties make NSs natural laboratories for studying the equation of state (EOS) of dense matter and physics under extreme conditions.
   
    As the density increases, new degrees of freedom may appear in the core of an NS, such as $\Delta$ resonances \cite{Glendenning1991,Drago2014a,Drago2014b,Zhu2016,Li2018,Thapa2022}, meson condensation \cite{Barshay1973,Baym1973,Pandharipande1995,Glendenning1999,Li2006}, hyperons \cite{Ambartsumyan1960,Glendenning1985,Schaffner1996,Shen2002,Weber2005,Weissenborn2012,Katayama2015,Huang2022a,Huang2024}, and quarks \cite{Baym1979,Celik1980,Glendenning1992,Satz1998}. The appearance of strange matter, such as hyperons, softens the EOS and reduces the maximum mass of NSs, giving rise to the well-known hyperon puzzle \cite{Burgio2021}. Quantum chromodynamics (QCD) predicts that, once the density exceeds several times the nuclear saturation density, hadronic matter may become deconfined and transform into quark matter, eventually forming a quark–gluon plasma \cite{Karsch2000,Karsch2008,Borsanyi2014,Bazavov2014}. The possible presence of quark matter inside NSs has therefore attracted significant attention and motivated extensive discussion 
    \cite{Baym1976,Steiner2000,Buballa2005,Oertel2017,annala2020,annala2023}.

   Several approaches have been proposed to describe the transition from hadronic matter to quark matter, including the Maxwell construction \cite{Glendenning1992}, the Gibbs construction \cite{Glendenning2001}, and the crossover model \cite{Kojo2015}. In the Maxwell construction, charge neutrality is imposed locally, and this approach applies when the surface tension between the two phases is very large. In contrast, the Gibbs construction enforces global charge neutrality and is applicable when the surface tension vanishes \cite{Constantinou2023}. For finite surface tension, the system may form nonuniform mixed phases with pasta-like geometries, commonly referred to as quark pasta, which can modify the properties of the transition region \cite{Ju2021}. When the hadronic and quark phases cannot simultaneously satisfy the equalities of pressure and chemical potential, the crossover model provides an alternative description by introducing a smooth interpolation. Typical interpolation schemes include parametrizing the pressure or energy density as a function of baryon number density \cite{Masuda2013a,Masuda2013b}, expressing the pressure as a function of energy density \cite{Hell2014}, or expanding the pressure as a polynomial in the chemical potential \cite{Kojo2015}. More recently, Gaussian process regression has also been applied to construct crossover EOSs 
   \cite{landry2019,essick2020,Huang2022}.
    
   At low densities, the EOS of NS matter can be constrained by the properties of finite nuclei and by chiral effective field theory around the nuclear saturation density \cite{Tews2013,Drischler2019}. However, the EOS of the outer core still carries large uncertainties, since terrestrial experiments cannot reach such high densities or reproduce the neutron-rich and $\beta$-equilibrated conditions of NS matter. At very high densities, above $40n_0$, perturbative QCD (pQCD) becomes reliable \cite{Gorda2018,Gorda2021}. Although this density range lies far beyond that realized in NS interiors, pQCD can still provide meaningful constraints on the EOS. Komoltsev and Kurkela showed that, by imposing thermodynamic stability and causality, pQCD constraints can be consistently propagated to lower-density nuclear matter \cite{Komoltsev2022}.
    
    In recent years, several studies have examined how pQCD can be used to constrain the EOS of NSs by enforcing thermodynamic stability, causality, and consistent matching between low- and high-density regimes  
\cite{Somasundaram2023,Albino2024,Malik2024,Zhou2025,Gao2025,komoltsev2024,finch2025,gorda2025,semposki2026}.
    At the same time, it is well established that quark matter forms Cooper pairs at sufficiently low temperatures and enters a color-superconducting phase, with energy gaps on all or part of the Fermi surfaces \cite{Bertrand1977,Bailin1984,Mark1998}. This phase introduces nonperturbative contributions to the EOS. The pQCD description can be consistently combined with the effects of color superconductivity \cite{Kurkela2024}. The color-superconducting phase should therefore be included when constraining the EOS in the high-density region.

    Over the past decade, advances in observational techniques have led to significant progress in our understanding of NS properties. Precise mass measurements of PSR J0348+0432 \cite{Antoniadis2013} and PSR J0740+6620 \cite{Cromartie2020} have confirmed that NS masses can exceed $2M_\odot$. The binary NS merger GW170817 has provided important constraints on the tidal deformability of NSs \cite{Hartley2017,Abbott2017}. In addition, intermediate-mass NSs, such as the recently reported PSR J0614–3329 with a radius of $R = 10.29^{+1.01}_{-0.86}~\mathrm{km}$ and a mass of $M = 1.44^{+0.06}_{-0.07}M_\odot$, impose even tighter constraints on the EOS \cite{Mauviard2025}. The coexistence of small radii at intermediate masses and large maximum masses requires the EOS to stiffen rapidly at higher densities. This behavior leaves a possibility of a transition from hadronic matter to quark matter in NS interiors.
    
    In this work, we further constrain the crossover EOS by combining recent astronomical observations with pQCD constraints. We describe the hadronic phase using relativistic mean-field models and the quark phase using the Nambu–Jona-Lasinio (NJL) model. The effects of pQCD and color superconductivity are incorporated at high densities. The theoretical framework, including the hadronic EOS, the NJL model, and the pQCD constraints, is introduced in Sec. II. The results and discussions are presented in Sec. III. Our conclusions are summarized in Sec. IV.
	
	\section{THEORETICAL FRAMEWORK}
    \subsection{Hadronic Matter EOS}
	Three relativistic mean-field (RMF) models are employed to describe the hadronic EOS, all of which can reproduce the properties of finite nuclei. These include the nonlinear RMF model (NLRMF) \cite{Walecka1974,Muller1996,Horowitz2001,Bao2014}, the density-dependent RMF model (DDRMF) \cite{Brockmann1992,Nikifmmode2002}, and the density-dependent point-coupling model (DDPC) \cite{Nikolaus1992,Burvenich2002}. In the following, we briefly summarize the theoretical framework of these models and present the expressions for the energy density and pressure used in this work. 
	
	In the NLRMF model, nucleons interact through the exchange of mesons, including the scalar–isoscalar $\sigma$ meson, the vector–isoscalar $\omega$ meson, and the vector–isovector $\rho$ meson. The interaction between the $\omega$ and $\rho$ mesons is also included. The Lagrangian density of the NLRMF model is given by
	\begin{align}
	\mathcal{L}_{\rm NLRMF}=& \sum_{N=n,p}\bar{\psi}_N\left[i\gamma^{\mu}\partial_{\mu}-\left(M_N-g_{\sigma}\sigma\right)-\gamma^{\mu}\left(g_{\omega}\omega_{\mu}+\frac{g_{\rho}}{2}\vec{\tau}\vec{\rho_\mu}\right)\right]\psi_N \nonumber\\
	&+\frac{1}{2}\partial^{\mu}\sigma\partial_{\mu}\sigma-\frac{1}{2}m_{\sigma}^2\sigma^2-\frac{1}{3}g_{2}\sigma^3
	-\frac{1}{4}g_{3}\sigma^4 \nonumber\\
	&-\frac{1}{4}W^{\mu\nu}W_{\mu\nu}+\frac{1}{2}m_{\omega}^2\omega^{\mu}\omega_{\mu}+\frac{1}{4}c_3\left(\omega^{\mu}\omega_{\mu}\right)^2 \nonumber\\
	&-\frac{1}{4}\vec{R}^{\mu\nu}\vec{R}_{\mu\nu}{+}\frac{1}{2}m_{\rho}^2\vec{\rho}^{\mu}\vec{\rho}_{\mu}
	+\Lambda_{\rm v}\left(g_{\omega}^2\omega^{\mu}\omega_{\mu}\right)\left(g_{\rho}^2\vec{\rho}^{\mu}\vec{\rho}_{\mu}\right),
	\end{align}
    where $W_{\mu\nu}$ and $\vec{R}_{\mu\nu}$ are the antisymmetric field tensors of the $\omega$ and $\rho$ mesons, respectively. In the mean-field approximation, all meson fields are treated as classical fields. For uniform matter, the total energy density and pressure of hadronic matter can be derived from the energy–momentum tensor and are expressed as,
    \begin{align}
    	\mathcal{E}_{\rm HP}^{\rm NLRMF}=&\frac{1}{2}m_{\sigma}^2\sigma^2+\frac{1}{3}g_{2}\sigma^3+\frac{1}{4}g_{3}\sigma^4\nonumber+\frac{1}{2}m_{\omega}^2\omega^2+\frac{3}{4}c_{3}\omega^4\\&+\frac{1}{2}m_{\rho}^2\rho^2+3\Lambda_{\rm v}\left(g_{\omega}^2\omega^{2}\right)\left(g_{\rho}^2\rho^{2}\right)\nonumber+\mathcal{E}_{\rm kin}^p+\mathcal{E}_{\rm kin}^n, \\ 
    	P_{\rm HP}^{\rm NLRMF}=& -\frac{1}{2}m_{\sigma}^2\sigma^2-\frac{1}{3}g_{2}\sigma^3-\frac{1}{4}g_{3}\sigma^4\nonumber+\frac{1}{2}m_{\omega}^2\omega^2+\frac{1}{4}c_{3}\omega^4\\&+\frac{1}{2}m_{\rho}^2\rho^2+\Lambda_{\rm v}\left(g_{\omega}^2\omega^{2}\right)\left(g_{\rho}^2\rho^{2}\right)+P_{\rm kin}^p+P_{\rm kin}^n.
	\end{align} 

    In the NLRMF model, the energy density and pressure of hadronic matter consist of contributions from the meson mean fields and from the kinetic terms of nucleons. The effective mass of nucleons is generated by the scalar $\sigma$ field, while the vector fields $\omega$ and $\rho$ provide repulsive interactions. The kinetic energy density and pressure are obtained by integrating over the Fermi seas of protons and neutrons,
    \begin{align}
        \mathcal{E}_{\rm kin}^i&=\frac{\gamma}{2\pi^2}\int_{0}^{k_{Fi}}k^2\sqrt{k^2+{M_i^{*}}^{2}}dk=\frac{\gamma}{16\pi^2}\left[k_{Fi}E_{Fi}^{*}\left(2k_{Fi}^2+{M_i^{*}}^2\right)+{M_i^{*}}^4{\rm ln}\frac{M_i^{*}}{k_{Fi}+E_{Fi}^{*}}\right], \nonumber\\
        P_{\rm kin}^i&=\frac{\gamma}{6\pi^2}\int_{0}^{k_{Fi}}\frac{k^4 dk}{\sqrt{k^2+{M_i^{*}}^{2}}}
        =\frac{\gamma}{48\pi^2}\left[k_{Fi}\left(2k_{Fi}^2-3{M_i^{*}}^2\right)E_{Fi}^{*}+3{M_i^{*}}^{4}{\rm ln}\frac{k_{Fi}+E_{Fi}^{*}}{M_i^{*}}\right],
    \end{align}
	where $\gamma = 2$ denotes the spin degeneracy factor.

    In the DDRMF model, the nonlinear meson self-interaction terms of the NLRMF model are replaced by density-dependent meson–nucleon coupling constants. The Lagrangian density of the DDRMF model is given by
    \begin{align}
        \mathcal{L}_{\rm DDRMF}
        =&\sum_{N=n,p}\overline{\psi}_ N\left[\gamma^{\mu}\left(i\partial_{\mu}-\Gamma_{\omega N}(n_ B)\omega_{\mu}-\frac{\Gamma_{\rho N}(n_B)}{2}\vec{\rho}_{\mu}\vec{\tau}\right) \right. \notag\\ 
        &\phantom{\bigg[}-\left(M_ N-\Gamma_{\sigma N}(n_B)\sigma-\Gamma_{\delta N}(n_{B})\vec{\delta}\vec{\tau}  \right)\bigg]\psi_N \nonumber\\
        &+\frac{1}{2}\left(\partial^{\mu}\sigma\partial_{\mu}\sigma-m_{\sigma}^2\sigma^2\right)
        +\frac{1}{2}\left(\partial^{\mu}\vec{\delta}\partial_{\mu}\vec{\delta}-m_{\delta}^2\vec{\delta}^2\right)\nonumber\\
        &-\frac{1}{4}W^{\mu\nu}W_{\mu\nu}+\frac{1}{2}m_{\omega}^2\omega_{\mu}\omega^{\mu}-\frac{1}{4}\vec{R}^{\mu\nu}\vec{R}_{\mu\nu}+\frac{1}{2}m_{\rho}^2\vec{\rho}_{\mu}\vec{\rho}^{\mu},
    \end{align} 
    where the coupling constants depend on the vector density. In this framework, rearrangement terms $\Sigma_{R}$ arise from the density dependence of the couplings and are essential for maintaining thermodynamic consistency. The resulting expressions for the energy density and pressure of hadronic matter are shown
    \begin{align}
        \mathcal{E}_{\rm HP}^{\rm DDRMF}=&\frac{1}{2}m_{\sigma}^2\sigma^2+\frac{1}{2}m_{\delta}^2\delta^2-\frac{1}{2}m_{\omega}^2\omega^2-\frac{1}{2}m_{\rho}^2\rho^2 +\Gamma_{\omega  N}(n_B)\omega n_B \notag\\ &+\frac{\Gamma_{\rho N}(n_B)}{2}\rho n_{B3}+\mathcal{E}_{\rm kin}^p+\mathcal{E}_{\rm kin}^n,\notag\\
        P_{\rm HP}^{\rm DDRMF}=&n_N\Sigma_{R}(n_B)-\frac{1}{2}m_{\sigma}^2\sigma^2-\frac{1}{2}m_{\delta}^2\delta^2+\frac{1}{2}m_{\omega}^2\omega^2
        \notag\\&+\frac{1}{2}m_{\rho}^2\rho^2+P_{\rm kin}^p+P_{\rm kin}^n.
    \end{align} 

    In the DDPC model, the meson exchange interactions are replaced by zero-range point-coupling terms between nucleons. Each interaction channel is represented by a local four-fermion interaction, with density-dependent coupling constants. The Lagrangian density of the DDPC model is given by
    \begin{align}
        \mathcal{L}_{\mathrm{DDPC}}= &\sum_{N=n,p} \overline{\psi}_ N\left(i\gamma^{\mu}\partial_{\mu}-M_N\right) \psi_N \nonumber\\
        & -\frac{1}{2} G_{S}(n_B)(\overline{\psi}_ N \psi_N)(\overline{\psi}_ N \psi_N)-\frac{1}{2} G_{V}(n_B)\left(\overline{\psi}_ N \gamma_{\mu} \psi_N\right)\left(\overline{\psi}_ N \gamma^{\mu} \psi_N\right) \nonumber\\
        & -\frac{1}{2} G_{T S}(n_B)(\overline{\psi}_ N \vec{\tau} \psi_N)(\overline{\psi}_ N \vec{\tau} \psi_N)-\frac{1}{2} G_{T V}(n_B)\left(\overline{\psi}_ N \vec{\tau} \gamma_{\mu} \psi_N\right)\left(\overline{\psi}_ N \vec{\tau} \gamma^{\mu} \psi_N\right) \nonumber\\
        & -\frac{1}{2} D_{S}(n_B)\left(\partial_{\mu} \overline{\psi}_ N \psi_N\right)\left(\partial^{\mu} \overline{\psi}_ N \psi_N\right).
    \end{align}
   
   In this approach, both scalar and vector interactions, as well as their isovector components, are included. The energy density and pressure of hadronic matter are obtained in a straightforward manner and are derived as
    \begin{align}
        \mathcal{E}_{\rm HP}^{\rm DDPC}= & \mathcal{E}_{\rm kin}^p+\mathcal{E}_{\rm kin}^n-\frac{1}{2} G_{S} n_{s}^{2}-\frac{1}{2} G_{T S} n_{s 3}^{2}+\frac{1}{2} G_{V} n_B^{2}+\frac{1}{2} G_{T V} n_{B3}^{2}, \nonumber\\
        P_{\rm HP}^{\rm DDPC}= & P_{\rm kin }^{n}+P_{\rm kin }^{p}+\frac{1}{2} G_{S} n_{s}^{2}+\frac{1}{2} G_{T S} n_{s 3}^{2}+\frac{1}{2} G_{V} n_B^{2}+\frac{1}{2} G_{T V} n_{B3}^{2} \nonumber\\
        & +\frac{1}{2} \frac{\partial G_{S}}{\partial n} n_{s}^{2} n_B+\frac{1}{2} \frac{\partial G_{T S}}{\partial n} n_{s 3}^{2} n_B+\frac{1}{2} \frac{\partial G_{V}}{\partial n} n_B^{3}+\frac{1}{2} \frac{\partial G_{T V}}{\partial n} n_{B3}^{2} n_B.
    \end{align}

     In these models, the scalar density $n_s$, baryon density $n_B$, and their isospin components are defined through the expectation values of the nucleon fields.
    \begin{align}
        n_s=&\langle \overline{\psi} \psi\rangle=n_{sp}+n_{sn}, \nonumber\\
        n_{s3}=&\langle \overline{\psi} \tau_3 \psi\rangle=n_{sp}-n_{sn}, \nonumber\\
        n_B=&\langle \psi^\dagger \psi\rangle=n_{Bp}+n_{Bn}, \nonumber\\
        n_{B3}=&\langle \psi^\dagger \tau_3 \psi\rangle=n_{Bp}-n_{Bn}.     
    \end{align}  
    
    For the hadronic EOS, we adopt three representative parameter sets, IUFSU \cite{Fattoyev2010a,Fattoyev2010b} for the NLRMF model, DDVTD \cite{Type2020}for the DDRMF model, and DDPC1 \cite{Nikifmmode2008}  for the DDPC model. These parameter sets are known to reproduce the properties of finite nuclei and provide reasonable descriptions of nuclear matter over a wide density range. They are therefore suitable for exploring the hadronic phase of NS matter in this work.
    
    \subsection{Quark Matter EOS and Crossover}
    To describe quark matter, we employ the three-flavor NJL model, which captures key features of low-energy QCD, such as dynamical chiral symmetry breaking \cite{Blaschke2005,Buballa2005,Ruester2005}.  In this framework, the NJL model provides an effective description of quark matter in the density range relevant for the crossover from hadronic matter. The Hatsuda–Kunihiro parameter set is adopted, where  $G_s\Lambda^2=1.835,~K\Lambda^5=9.29$, with $\Lambda=631.4~\rm{MeV}$. The Lagrangian of the three-flavor NJL model is:
    \begin{align}
    \mathcal{L}_{\mathrm{NJL}}=&\overline{q}(i\gamma^{\mu}\partial_{\mu}-m) q+G_{s} \sum_{i=0}^{8}\left[\left(\overline{q} \tau_{i} q\right)^{2}+\left(\overline{q} i \gamma_{5} \tau_{i} q\right)^{2}\right] -G_{v}\left(\overline{q} \gamma^{\mu} q\right)^{2} \nonumber \\
    &+H\sum_{A, A^{\prime}=2,5,7}\left[\left(\overline{q} i \gamma_{5} \tau_{A} \lambda_{A^{\prime}} C \overline{q}^{T}\right)\left(q^{T} C i \gamma_{5} \tau_{A} \lambda_{A^{\prime}} q\right)+\left(\overline{q} \tau_{A} \lambda_{A^{\prime}} C \overline{q}^{T}\right)\left(q^{T} C \tau_{A} \lambda_{A^{\prime}} q\right)\right] \nonumber \\
    &-K\left\{\operatorname{det} \overline{q}\left(1+\gamma_{5}\right) q+\operatorname{det} \overline{q}\left(1-\gamma_{5}\right) q\right\}.
    \end{align}
    It includes the scalar four-quark interaction with coupling constant $G_s$, which drives chiral symmetry breaking, and the Kobayashi–Maskawa–’t Hooft (KMT) determinant interaction with coupling $K$, which accounts for the $U(1)_A$ anomaly. A vector interaction term with coupling $G_v$ is introduced to describe repulsive interactions among quarks. In addition, diquark interaction terms with coupling $H$ are included to account for attractive correlations in the color-superconducting channel. The coupling constants $G_v$ and $H$ are not well constrained and play an important role in determining the stiffness of the quark-matter EOS.
    
    The pressure of the quark phase from the NJL model is generated from the thermodynamic potential,
    \begin{equation}
        P_{\rm NJL}=2 \sum_{i=1}^{18} \int^{\Lambda} \frac{d^{3} \mathbf{p}}{(2 \pi)^{3}} \frac{\epsilon_{i}}{2}-\sum_{i=u, d, s}\left(2 G_s \sigma_{i}^{2}+H d_{i}^{2}\right)+4 K \sigma_{u} \sigma_{d} \sigma_{s}+G_{v} n_{q}^{2} ,
    \end{equation}
    where $\sigma$, $d$, and $n_q$ represent the chiral condensates, quark condensates, and quark number density, respectively. $\epsilon_{i}$ are the energy eigenvalues obtained from the inverse propagator in the Nambu-Gorkov bases,
    \begin{equation}
        S^{-1}(k)=\left(\begin{array}{lc}
        \gamma_{\mu} k^{\mu}-\hat{M}+\gamma^{0} \hat{\mu} & \gamma_{5} \sum_{i} \Delta_{i} R_{i} \\
        -\gamma_{5} \sum_{i} \Delta_{i}^{*} R_{i} & \gamma_{\mu} k^{\mu}-\hat{M}-\gamma^{0} \hat{\mu}
        \end{array}\right),
    \end{equation}
    where $M_i$ are the constituent masses of $u$, $d$, $s$ quarks and $\Delta_{1,2,3}$ are the gap energies. $\mu_3$, $\mu_8$ are the color chemical potentials, $Q=\rm{diag}(2/3,-1/3,-1/3)$ is the charge matrix in flavor space,
    \begin{align}
    M_{i} & =m_{i}-4 G_s \sigma_{i}+K\left|\epsilon_{i j k}\right| \sigma_{j} \sigma_{k}, \nonumber\\
    \Delta_{i} & =-2 H d_{i}, \nonumber\\
    \hat{\mu} & =\mu_{q}-2 G_{v} n_{q}+\mu_{3} \lambda_{3}+\mu_{8} \lambda_{8}+\mu_{Q} Q,\nonumber \\
    \left(R_{1}, R_{2}, R_{3}\right) & =\left(\tau_{7} \lambda_{7}, \tau_{5} \lambda_{5}, \tau_{2} \lambda_{2}\right).
    \end{align}
    The mean fields are determined from the gap equations,
    \begin{equation}
        \frac{\partial P_{\rm NJL}}{\partial \sigma_i} = \frac{\partial P_{\rm NJL}}{\partial d_i}=0.
    \end{equation}
    To enforce the conditions of local electromagnetic charge neutrality and color charge neutrality, three other equations to be satisfied,
    \begin{equation}
        \frac{\partial P_{\rm NJL}}{\partial \mu_3} = \frac{\partial P_{\rm NJL}}{\partial \mu_8}=\frac{\partial P_{\rm NJL}}{\partial \mu_Q}=0.
    \end{equation}
    The pressure and quark chemical potential can be rewritten by explicitly separating the contributions from the vector interaction. In this form, the effects of the vector coupling $G_v$ enter only through the quark number density,
    \begin{equation}
    \begin{aligned}
        P_{\rm NJL}=&P_{G_{v}=0}+G_{v} n_{q}^{2}, \\
        \mu_q=&\mu_{G_{v}=0} +2G_vn_q.     
    \end{aligned}
    \end{equation}
    This treatment allows the thermodynamic quantities to be obtained from the $G_v=0$ case and then extended to arbitrary values of $G_v$ in a straightforward manner. 
    
    When considering the constraints of pQCD, the parameter space of the NJL model is greatly reduced and it is difficult to match astronomical constraints. Therefore, a chemical potential dependent NJL model was proposed \cite{Pinto2023}. To better address the thermodynamic consistency of the  NJL model with medium-induced contributions, the density-dependent NJL (DDNJL) model is derived, where the parameter $G_v$ depends on the vector density. In the mean-field approximation, the terms related to $G_v$ in the Lagrangian of the NJL model can be rewritten as:
    \begin{equation}
        \begin{aligned}
            G_v((q^\dagger q) )(\bar{q}\gamma^\mu q)^2
            {\simeq} &\left( G_v(\langle q^\dagger q\rangle) + \frac{\partial G_v}{\partial n_q}\left (  (q^\dagger q)-\langle q^\dagger q\rangle\right )  \right)(\bar{q}\gamma^\mu q)^2  \\=&G_v(n_q)\left( 2\langle q^\dagger q\rangle(q^\dagger q) - \langle q^\dagger q\rangle^2\right)+\frac{\partial G_v}{\partial n_q} \left( (q^\dagger q)\langle q^\dagger q\rangle^2-\langle q^\dagger q\rangle^3  \right).
        \end{aligned}
    \end{equation}
    Therefore, the pressure and quark chemical potential are modified by the derivative of $G_v$ with respect to density, which reflect the influence  of the medium effect on the vector interaction, 
    \begin{equation}
    \begin{aligned}
      P_{\rm DDNJL}=&2 \sum_{i=1}^{18} \int^{\Lambda} \frac{d^{3} \mathbf{p}}{(2 \pi)^{3}} \frac{\epsilon_{i}}{2}-\sum_{i=u, d, s}\left(2 G_s \sigma_{i}^{2}+H d_{i}^{2}\right)\\&+4 K \sigma_{u} \sigma_{d} \sigma_{s}
      +G_{v} n_{q}^{2} +\frac{\partial G_v}{\partial n_q}n_q^3\\
      =&P_{G_v=0}+G_vn_q^2+\frac{\partial G_v}{\partial n_q}n_q^3,\\
      \hat{\mu}_{\rm DDNJL} =&\mu_{q}-2 G_{V} n_{q}-\frac{\partial G_v}{\partial n_q}n_q^2+\mu_{3} \lambda_{3}+\mu_{8} \lambda_{8}+\mu_{Q} Q ,\\
     \mu_q=&\mu_{G_{v}=0}+2G_vn_q+\frac{\partial G_v}{\partial n_q}n_q^2=\mu_{G_{v}=0}+2G_v^*n_q,\\
    \end{aligned}
    \end{equation}
    with
    \begin{equation}\label{gvn}
    G_v^*=G_v+\frac{1}{2}\frac{\partial G_v}{\partial n_q}n_q.
    \end{equation}
    which combines the original density-dependent coupling and the rearrangement contribution. This effective coupling $G_v^*$ governs the strength of the vector interaction in the quark chemical potential. It plays a key role in controlling the stiffness of the quark-matter EOS at different densities.

    In present unified EOS, the region with a baryon number density below $2n_0$ is the hadronic phase, and the region above $5n_0$ is the quark phase. The crossover region between the hadron phase to the quark phase is described by a smooth interpolation in the pressure-chemical potential space. The pressure is expanded as a fifth-order polynomial of the chemical potential,
    \begin{equation}
        P_I(\mu_B)=\sum_{i=0}^{5} C_i\mu_B^i,
    \end{equation}
    and the parameters $C_i$ are determined by the boundary conditions,
    \begin{equation}
    \begin{aligned}
        \left.\frac{\mathrm{d}^{n} P_{\mathrm{I}}}{\mathrm{~d} \mu_{B}^n}\right|_{\mu_{L}} & =\left.\frac{\mathrm{d}^{n} P_{\mathrm{_H}}}{\mathrm{~d} \mu_{B}^n}\right|_{\mu_{L}} \\
        \left.\frac{\mathrm{~d}^{n} P_{\mathrm{I}}}{\mathrm{~d} \mu_{B}^n}\right|_{\mu_{H}} & =\left.\frac{\mathrm{d}^{n} P_{\mathrm{Q}}}{\mathrm{~d} \mu_{B}^{n}}\right|_{\mu_{H}}, \quad(n=0,1,2),
    \end{aligned}
    \end{equation}
    where $\mu_{L}$ and $\mu_{H}$ are the chemical potential corresponding to $n_B=2.0n_0$ and $n_B=5.0n_0$. The pressure and its first and second derivatives with respect to the baryon chemical potential are required to match those of the hadronic EOS at $\mu_L$ and the quark EOS at $\mu_H$. These conditions ensure the continuity of the pressure, baryon density, and compressibility across the crossover region.
    
    \subsection{Constraints from pQCD}
    As a nonrenormalizable theory, the NJL model cannot absorb the cutoff parameters required to regularize divergent momentum integrals into renormalized quantities \cite{Klevansky1992}. The cutoff scale therefore limits the range of applicability of the model. This limitation appears as cutoff artifacts in NJL calculations.  {Due to cutoff artifacts, as the chemical potential approaches the cutoff scale, the pairing gap begins to decrease and eventually vanishes\cite{Farias2006,braguta2016,kogut2000}. When addressing this problem, a reliable approach is the renormalization group improved NJL model\cite{Gholami2025a}. However, in this work, we determine the valid regime of the NJL model by identifying the region where the ultraviolet cutoff has negligible effect.}

    To determine the upper limit of the effective range of the NJL model, we first perform calculations with $G_v = 0$. We identify the chemical potential at which the pairing gap $\Delta$ starts to decrease. The termination point of the effective chemical potential range is then defined as a value $60~\mathrm{MeV}$ below this chemical potential. The termination points, denoted by $\mu_T$, for arbitrary values of $G_v$ are obtained using the method described in the previous subsection.

     The approach in Ref. \cite{Komoltsev2022} is followed to impose the constraints of pQCD, and  thermodynamic stability and causality are used to determine whether the two endpoints $(n_L,\mu_L,P_L)$ and $(n_H,\mu_H,P_H)$ can be connected, where $\mu_L=\mu_T$, $\mu_H=2.6~\rm{GeV}$.
    \begin{equation}\label{prere}
        \Delta P_{min}\le P_H-P_L\le \Delta P_{max},
    \end{equation}
    where,
    \begin{equation}
    \begin{aligned}
        \Delta P_{min} &= \frac{1}{2} \frac{n_L}{\mu_L}(\mu_H^2-\mu_L^2), \\
        \Delta P_{max} &= \frac{1}{2} \frac{n_H}{\mu_H}(\mu_H^2-\mu_L^2). 
    \end{aligned}
    \end{equation}
    
    At high chemical potential, the matter in the NS is assumed to be in the color–flavor locked (CFL) phase. This phase preserves the symmetry under combined color–flavor rotations. The pressure and the baryon number density are given by,
    \begin{equation}
    \begin{aligned}
        P_H(\mu_H) &= P_{pQCD}(\mu_H)+P_{CFL}(\mu_H)=P_{pQCD}(\mu_H)+\frac{1}{3\pi^2}\Delta^2\mu_H^2, \\
        n_H(\mu_H) &= n_{pQCD}(\mu_H)+n_{CFL}(\mu_H)=n_{pQCD}(\mu_H)+\frac{2}{3\pi^2}\Delta^2\mu_H. 
    \end{aligned}
    \end{equation}
    where the second term on the right-hand side represents the contribution of CFL colour-superconducting \cite{Alford1999,Alford2008}. 
    
    The pQCD results are obtained from the weak-coupling expansion of QCD up to next-to-next-to-leading order (N$^2$LO). The strong coupling constant $\alpha_s$ is evaluated at the two-loop level \cite{Kurkela2010,Gorda2023a,Vuorinen2003}. The pressure of unpaired quark matter with $N_f = 3$ massless flavors is given by
    \begin{align}
    P_{pQCD}=&N_{f} \frac{\left(\mu_{B} / 3\right)^{4}}{12 \pi^{2}}\left\{1-\frac{2}{\pi} \alpha_{s}-\frac{N_{f}}{\pi^{2}} \alpha_{s}^{2} \ln \alpha_{s}-\frac{\alpha_{s}^{2}}{\pi^{2}}\right. \\
    &\left.\times\left[c_{1}+N_{f}\left(\ln \frac{N_{f}}{\pi}-c_{2}\right)+\left(11-\frac{2}{3} N_{f}\right) \ln X\right]\right\}, \\
    \alpha_s(\bar\Lambda)=&\frac{4\pi}{\beta_0 L}\left[1-\frac{2\beta_1}{\beta_0^2}\frac{\mathrm{ln} L}{L} \right],
    \end{align}
    with 
    \begin{align}
    L=&\ln(\bar\Lambda^2/\Lambda^2_{\overline{MS}}),\ \bar\Lambda = 2X\mu_B/3,\\
    \beta_0=&11-\frac{2}{3}N_f,\ \beta_1=51-\frac{19}{3}N_f,
    \end{align}
    where $c_1=18 - 11\mathrm{ln}2$ and $c_2=0.535832$, are constants. The quantity $\bar{\Lambda}$ denotes the renormalization scale, and $X$ is the corresponding dimensionless scale parameter. The QCD scale $\Lambda_{\overline{\mathrm{MS}}}$ is fixed by the condition $\alpha_s(2~\mathrm{GeV}) = 0.2994$.
    
     {We assume that the pairing gap follows a power-law dependence on the chemical potential as shown in Ref. \cite{geissel2025},
    \begin{equation}
        \Delta_{\rm CFL} = \Delta_{\rm CFL}^* \left( \frac{\mu_B}{\mu_B^*} \right)^{\alpha}
    \end{equation}
    where $\mu_B^*$ is the reference chemical potential and $\Delta_{\rm CFL}^*$ is the corresponding gap. $\alpha$ is a constant exponent, where $\alpha\approx -0.23$ in the weak-coupling case and $\alpha\approx 0.45$ in the functional renormalization group approach \cite{son1999,geissel2025}. These values well approximate the gap derived from both theories. Taking $2.0~\rm GeV$ as the reference chemical potential\cite{geissel2024,braun2024}, the gap at point H in the weak-coupling case is approximately $\Delta_{H}\approx 0.94\Delta_{2.0~\rm GeV}$. In fact, for large $G_v$, $\mu_L$ already exceeds $2.0~\rm GeV$. If a more conservative gap range of $[0.9,~1.1]\Delta_L$ is chosen, it will lead to a deviation of about $0.05G_s$ from the case with $\Delta_H=\Delta_L$. Therefore, we select the same gap value for two points. In the NJL model, the gaps $\Delta_1$ and $\Delta_3$ are generally not equal. We therefore define an effective pairing gap using the root-mean-square form, $\Delta = \sqrt{\frac{1}{3}\sum_i \Delta_i^2}$ \cite{Gholami2025}.

    We adopt the scale-averaging interpretation to quantify the pQCD uncertainties \cite{Gorda2023b}. The dimensionless renormalization scale $X$ is assumed to follow a logarithmically uniform distribution in the interval $[1/2,~2]$. This choice allows us to construct a likelihood function that quantifies whether a given EOS is consistent with the pQCD constraints. The resulting pQCD likelihood function is
    \begin{equation}\label{prf}
        PR=\int_{\ln(1/2)}^{\ln(2)} d(\ln X)\times \mathbb{1}_{[0,1]} (I_{pQCD}(X,EOS)),
    \end{equation}
    where $\mathbb{1}_{[0,~1]} (I_{pQCD}(X,EOS))$ is an indicator function that determines whether the EOS satisfies the pQCD constraints for a given value of $X$.
    
	\section{RESULTS AND DISCUSSIONS}\label{III}

    \begin{figure}[htbp]
        \centering
        \includegraphics[width=0.45\textwidth]{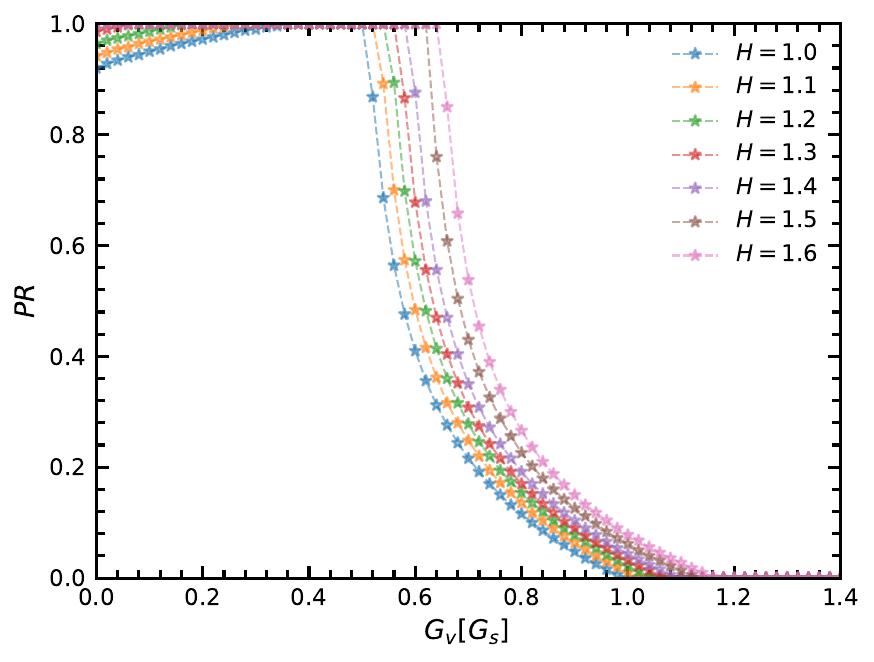}
        \caption{The pQCD likelihood function as a function of the vector and diquark coupling constants, $G_v$ and $H$ in the NJL model.}
        \label{fig1}
    \end{figure}
    The pQCD constraints on the vector and diquark coupling constants, $G_v$ and $H$ with the pQCD likelihood functions from Eq.~(\ref{prf})  are illustrated in Fig.~\ref{fig1}. In principle, when the value of $PR$ is located in $[0,~1]$, the EOS from NJL at high density should satisfy the pQCD constraints. Therefore, the magnitude of $G_v$ should be less than $1.2G_s$ at $H=1.6G_s$. If $G_v$ was too small, the EOS will be softer. It cannot be matched to the pQCD results at small values of $X$ within the thermodynamically allowed range. Therefore, the $PR<1$ for small $G_v$. As $G_v$ increases, the NJL EOS becomes stiffer, allowing the matching condition to be satisfied for all $X$ in the interval $[0.5~,2]$. For sufficiently large $G_v$, however, the EOS becomes overly stiff, and the pQCD results at large $X$ can no longer be connected to the effective endpoint of the NJL model. Meanwhile, as the diquark coupling $H$ decreases, the pQCD constraints become more restrictive, leading to a further reduction in the allowed range of $G_v$. It is {less than $1.0G_s$} at $H=1.0G_s$.

     \begin{figure}[htbp]
        \centering
        \includegraphics[width=0.7\textwidth]{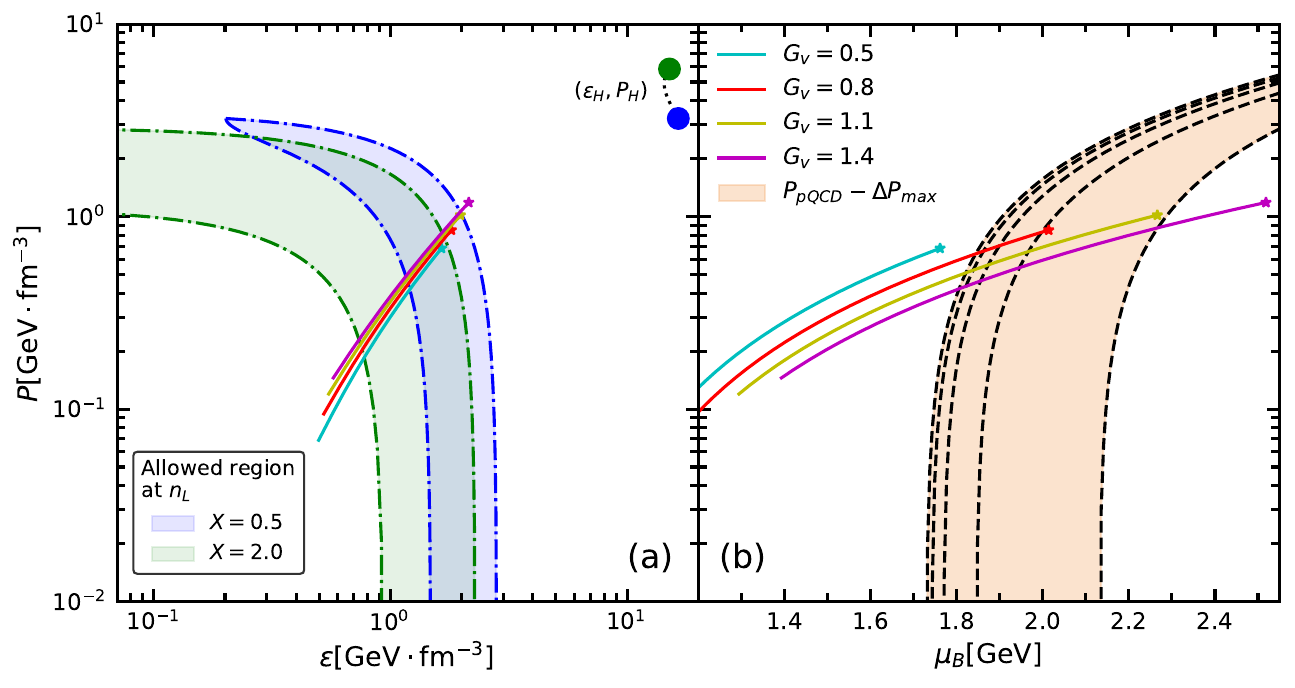}
        \caption{(Left panel)  EOSs from NJL models with different $G_v$ at $H=1.5G_s$ compared to the pQCD likelihoods in $p-\varepsilon$ plane with  $n_L=1.32$ fm$^{-3}$ and $X=0.5,~2.0$. The endpoints are marked as stars.  (Right panel) These EOSs also compared to the lower limits of the pressure  $P_R-\Delta P_{max}$ in the $p-\mu$ plane. }
        \label{fig2}
    \end{figure}
    
     Due to the ultraviolet cutoff, the NJL model is applicable only up to a finite chemical potential or baryon density, which defines an effective endpoint for matching to pQCD. {In this work, the endpoint is chosen at the chemical potential $60$ MeV below the point where the quark pairing gap starts to decrease for $G_v = 0$.} To avoid numerical instabilities, the chemical potential is fixed at $1340~\mathrm{MeV}$, which corresponds to a baryon density of $n_L = 1.32~\mathrm{fm}^{-3}$ and can be extrapolated to an arbitrary $G_v$.  
     
     In panel (a) of Fig.~\ref{fig2}, the EOSs obtained from the NJL model with different values of $G_v$ at $H = 1.5G_s$ are compared with the pQCD likelihood regions at $n_L = 1.32~\mathrm{fm}^{-3}$ and $X = 0.5$ and $2.0$ in the $p$–$\varepsilon$ plane. It is found that the endpoint of the EOS with $G_v = 1.4G_s$ lies outside the allowed shaded region and should therefore be excluded. In panel (b), the same EOSs are compared with the $P_R - \Delta P_{\max}$ constraint in the $p$–$\mu$ plane. From Eq.~(\ref{prere}), the condition $P_L > P_H - \Delta P_{\max}$ must be satisfied, with $P_H = P_{\mathrm{pQCD}} +P_{\rm CFL}$. Therefore, if an EOS is consistent with the pQCD constraints, the pressure from the NJL model should remain above the lower bound $P_{\mathrm{H}} - \Delta P_{\max}$. For $G_v = 1.4G_s$, the NJL pressure around $\mu = 2.2~\mathrm{GeV}$ already exceeds this lower bound. This result shows that the constraints illustrated in panels (a) and (b) are fully consistent with each other.

    The vector and diquark coupling constants, $G_v$ and $H$, are constrained not only by pQCD but also by NS observables and by the behavior of the speed of sound, $c_s^2$, in the hadron–quark crossover region. In the present work, the crossover density range is taken to be $2$–$5n_0$. For the crust region, we adopt results calculated with the IUFSU parameter set, which has a symmetry energy slope $L$ close to those of the models considered for uniform matter. Three hadronic outer-core EOSs based on RMF parametrizations, DDPC1, DDVTD, and IUFSU, are included in the analysis. The saturation properties of symmetric nuclear matter—including the saturation density $n_0$, energy per nucleon $E/A$, incompressibility $K$, skewness parameter $Q$, symmetry energy $E_{\rm sym}$, symmetry-energy slope $L$, and symmetry incompressibility $K_{\rm sym}$—obtained from these parameter sets are summarized in Table~\ref{table.sat}. It is found that the density dependence of the symmetry energy differs significantly among the three models, with the DDPC1 parameter set exhibiting a stiffer behaviour.
    \begin{table*}[htbp]
    	\footnotesize
    	\centering
    	\caption{Saturation properties of nuclear matter.}\label{table.sat}
    	\scalebox{0.8}{
    		\begin{tabular}{r|cccccccccccccc}
    			\hline\hline
    			&$n_{0}[\rm fm^{-3}]$ ~&$E/A[\rm MeV]$ ~&$K[\rm MeV]$ ~&$Q[\rm MeV]$  ~&$E_{\rm sym}[\rm MeV]$  ~&$L[\rm MeV]$ ~&$K_{\rm sym}[\rm MeV]$ ~&$Q_{\rm sym}[\rm MeV]$ \\
    			\hline 
    			DDPC1 ~&0.1519 ~&-16.06~&229.99~&-1118.23~&33.00 ~&70.12 ~& -108.00 ~& 201.11  \\
    			DDVTD  ~&0.1535 ~&-16.91~&239.34~&-762.46 ~&31.80 ~&42.58 ~&-117.39 ~& 873.05 \\ 
    			IUFSU ~&0.1547 ~&-16.40~&231.55~&-289.77 ~&31.31 ~&47.24 ~&28.642 ~& 369.13   \\
    			\hline\hline    
    	\end{tabular}}
    \end{table*}

    The diquark coupling constant $H$ is tightly constrained to be $H \simeq 1.5G_s$ by NS mass–radius measurements and by the behavior of $c_s^2$. For the DDPC1 and DDVTD parameter sets, the allowed range of $G_v$ is mainly determined by the causality condition $c_s^2 < 1$. This leads to $G_v \simeq 0.680$–$0.935G_s$ for DDPC1 and $G_v \simeq 0.689$–$1.017G_s$ for DDVTD. In contrast, for the IUFSU parameter set, the upper limit of $G_v$ is primarily constrained by pQCD. As a result, the allowed range is $G_v \simeq 0.812$–$1.1G_s$.
    
    Furthermore, the NJL model is now also  extended to the density-dependent vector coupling case in this work to better satisfy the pQCD constraints. It is functional form is 
    \begin{equation}
    	G_v(n_q)=\frac{a}{1+\exp(bn_q)}G_s,
    	\end{equation}  
    where $n_q$ is the quark number density, $a,~b$ are determined by the NS observables, $c^2_s$, and pQCD. {In Ref. \cite{Pinto2023}, $G_v$ was proposed as the chemical potential dependence. In this work, it is assumed as the quark number density-dependence, which was hinted by the density-dependent RMF model in hadron level. The present form will be convenient to calculate the arrangement term in Eq. (\ref{gvn}) with an analytical form. The two free parameters, $a~,b$ are chosen such that the speed of sound in the crossover region reaches its limit and satisfies the constraints from the pQCD likelihood function.} They are $a=2.20,~b=1/7$ for DDPC1, $a=2.48,~b=1/5$ for DDVTD set, and, $a=2.91,~b=1/5$ for IUFSU set, respectively. For convenient description, these three density-dependent quark vector coupling parameters are named as $G^1_v,~G^2_v$, and $G^3_v$.
    
  The relationships between pressure and energy density for the crossover EOSs, constructed from three types of RMF hadronic models and the NJL quark model, are presented in Fig.~\ref{fig3}. In this figure, constraints from neutron-star observables, the sound-speed condition, and pQCD are taken into account. The dashed {(dash-dotted)} lines indicate the crossover region spanning densities from $2$ to $5n_0$. The allowed shaded regions are bounded by different choices of the density-independent vector coupling constant $G_v$.
  
  The pure hadronic EOS based on the DDPC1 parameter set is extremely stiff.  At high densities, it is even stiffer than the pure quark-matter EOS obtained from the NJL model. In contrast, the EOSs derived from the DDVTD and IUFSU parameter sets are significantly softer. Furthermore, quark-matter EOSs generated with a density-dependent vector coupling $G_v(n_B)$ are softer than those obtained with a constant $G_v$. Consequently, in order to support massive neutron stars, the crossover EOS must become sufficiently stiff in the intermediate energy-density region.
  \begin{figure}[htbp]
        \centering
        \includegraphics[width=1\textwidth]{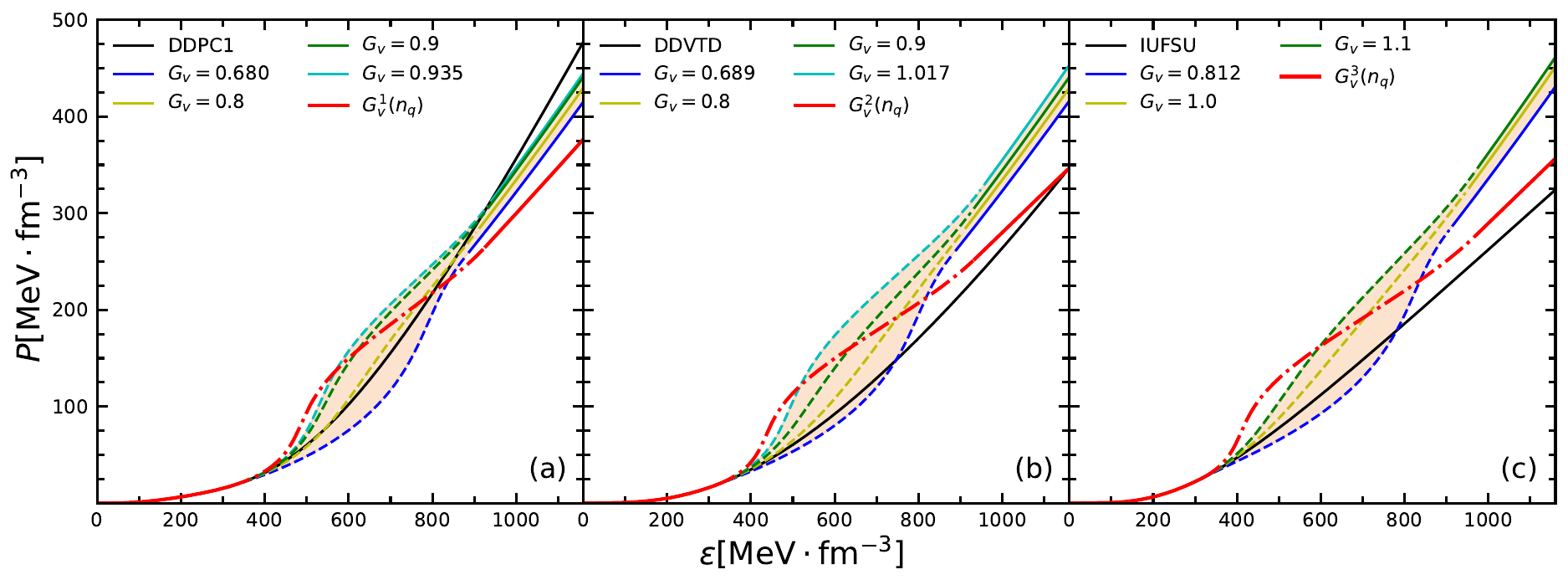}
        \caption{The pressure as a function of energy density of the crossover EOSs constructed by three RMF hadronic and NJL quark models.}
        \label{fig3}
    \end{figure}
    
The speed of sound of NS matter, $c_s^2$, as a function of baryon density is shown in Fig.~\ref{fig4} for the crossover EOSs displayed in Fig.~\ref{fig3}. A pronounced peak in the speed of sound appears in the crossover region. When the density-independent vector coupling is weak, this peak occurs at higher densities. In contrast, for stronger vector coupling, the peak shifts to lower densities. For the DDNJL models, $c_s^2$ rapidly reaches its maximum value and then decreases sharply. All of the present parameter sets for $G_v$ obey the causality limit $c_s^2 \le 1$.
    \begin{figure}[htbp]
        \centering
        \includegraphics[width=1\textwidth]{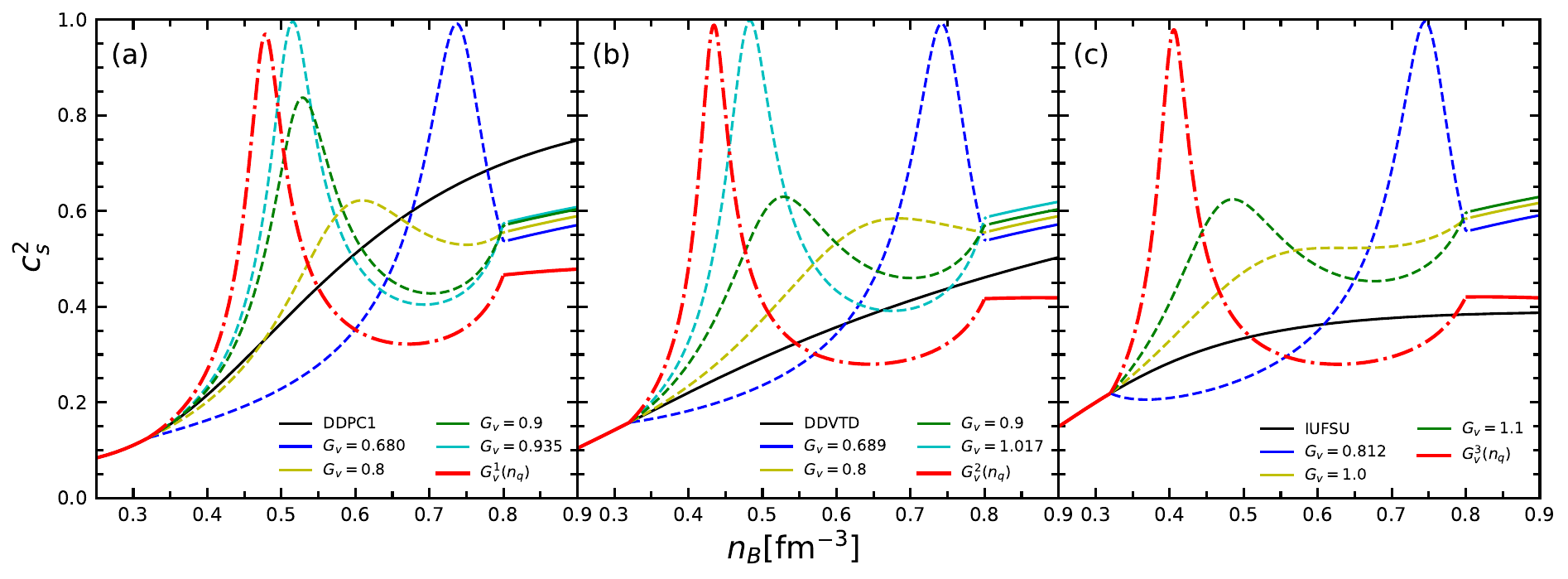}
        \caption{The square  sound speed as a function of baryon density of the crossover EOSs shown in Fig.~\ref{fig3}.}
        \label{fig4}
    \end{figure}

    \begin{figure}[htbp]
        \centering
        \includegraphics[width=1\textwidth]{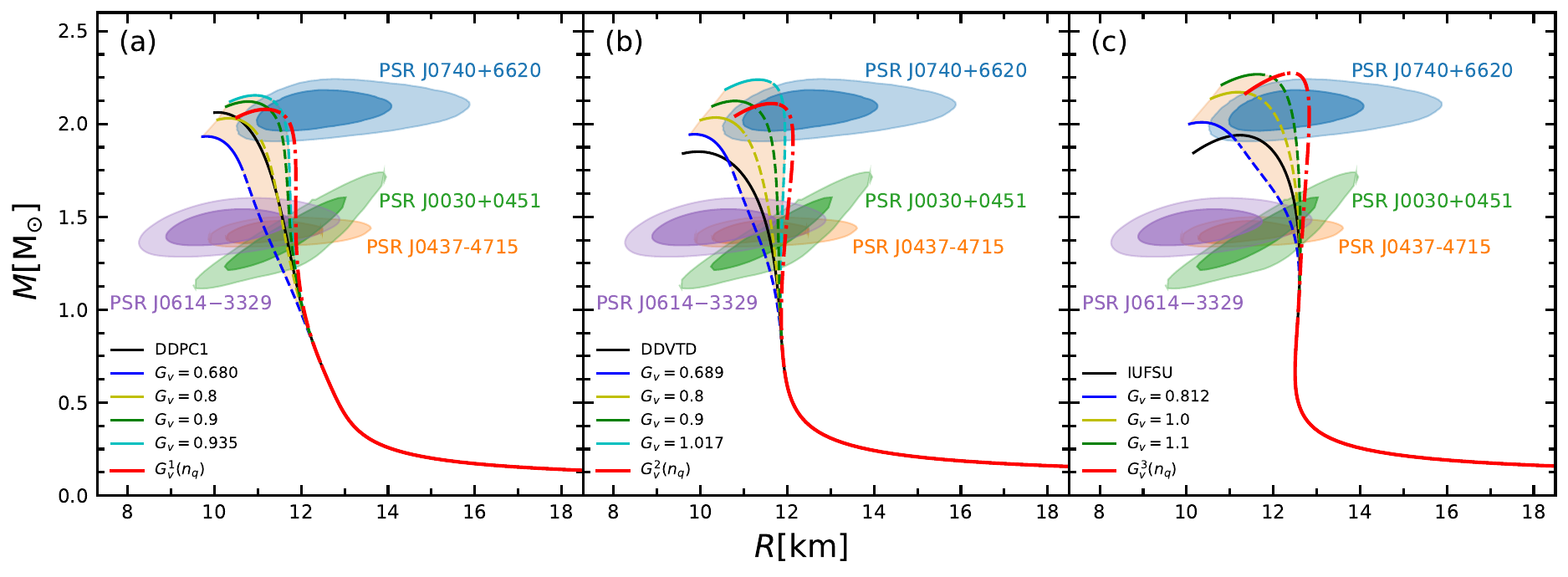}
        \caption{The mass-radius relations of NSs obtained with our constructed crossover EOSs and pure hadronic EOSs, together with observational constraints.}
        \label{fig5}
    \end{figure}

   The mass–radius relations of NSs obtained with our constructed crossover EOSs and pure hadronic EOSs are shown in Fig.~\ref{fig5}. Observational constraints from four NSs, PSR J0740+6620, PSR J0030+0451, PSR J0437–4715, and PSR J0614–3329, are also indicated. The results for pure hadronic matter are shown as solid curves. For the DDVTD and IUFSU parameter sets, the pure hadronic EOSs are too soft to reach the mass requirement of PSR J0740+6620.
   
   Compared with the pure hadronic case, most crossover EOSs can support massive NSs with maximum masses exceeding $2M_\odot$. The increase in the maximum mass is more pronounced for softer hadronic EOSs. For the DDVTD set, the maximum mass increases by nearly $20\%$. Recent observations of PSR J0614–3329 suggest that NSs should have relatively small radii at the canonical mass of $1.4M_\odot$. When the crossover occurs with a weak density-independent vector coupling in the NJL model, the radii of intermediate-mass NSs are reduced and become more consistent with the constraints from PSR J0614–3329. In contrast, in the DDNJL model, the radii around $1.4M_\odot$ are noticeably larger than those obtained with density-independent vector coupling. This difference arises because the corresponding crossover EOSs are stiffer, which also allows the formation of more massive NSs.
       
    \begin{figure}[htbp]
        \centering
        \includegraphics[width=1\textwidth]{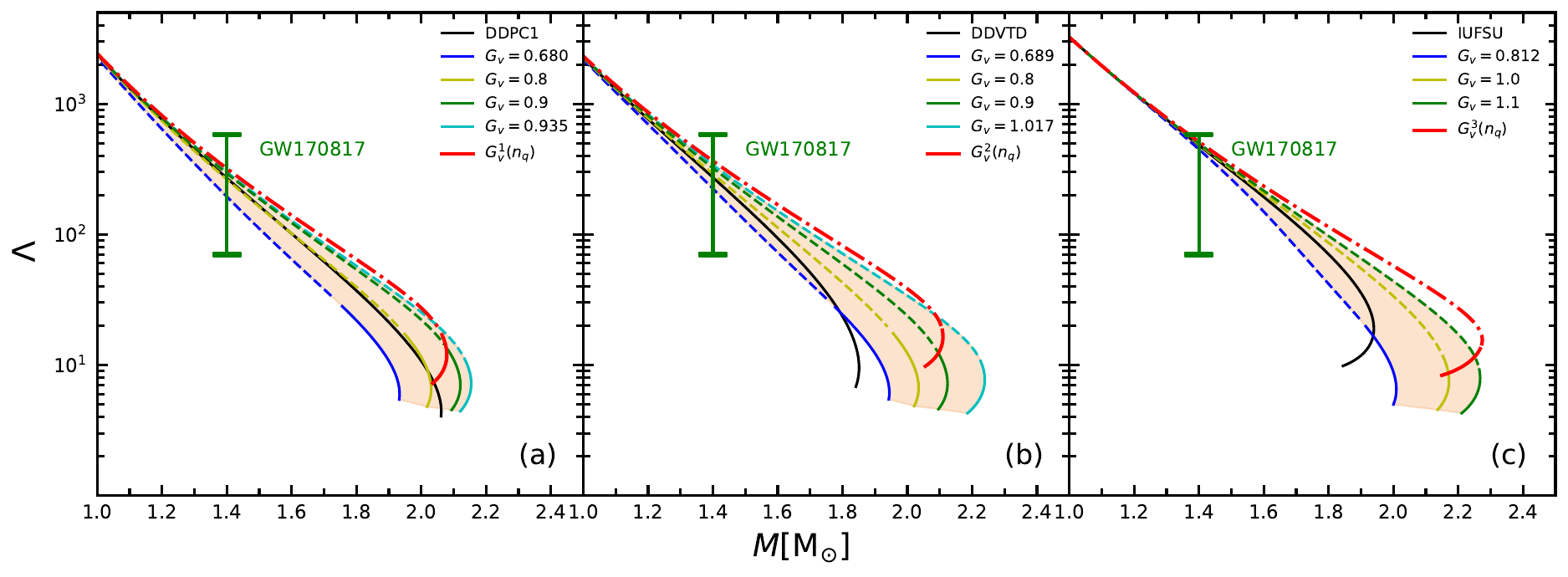}
        \caption{The tidal deformability as a function of NS mass with different crossover and pure hadronic EOSs.}
        \label{fig6}
    \end{figure}
    
    The dimensionless tidal deformability, $\Lambda$, is a new property of NS that quantifies how a compact object is deformed  under the action of an external gravitational field \cite{Hinderer2008,Hinderer2010} in the multi-messenger astronomy era. In Fig. \ref{fig6}, it is presented as a function of NS mass, and the constraint from the GW170817 event is also included. The tidal deformabilities at $1.4M_\odot$ from the pure hadronic EOSs generated by the three RMF sets can all satisfy the constraints of GW170817. When the quark phase is included, the magnitude of $\Lambda$ is obviously influenced by the vector coupling constants. The weak coupling can reduce $\Lambda$ and strong coupling will make it increase. They are in the ranges of $\Lambda$ as $193-300$, $223-339$, and $448-496$, with DDPC1, DDVTD, and IUFSU, respectively, with the density-independent vector coupling constant $G_v$. Once the density-dependence of $G_v$ is considered, the $\Lambda$ at $1.4M_\odot$ will be larger.

    \begin{figure}[htbp]
        \centering
        \includegraphics[width=1\textwidth]{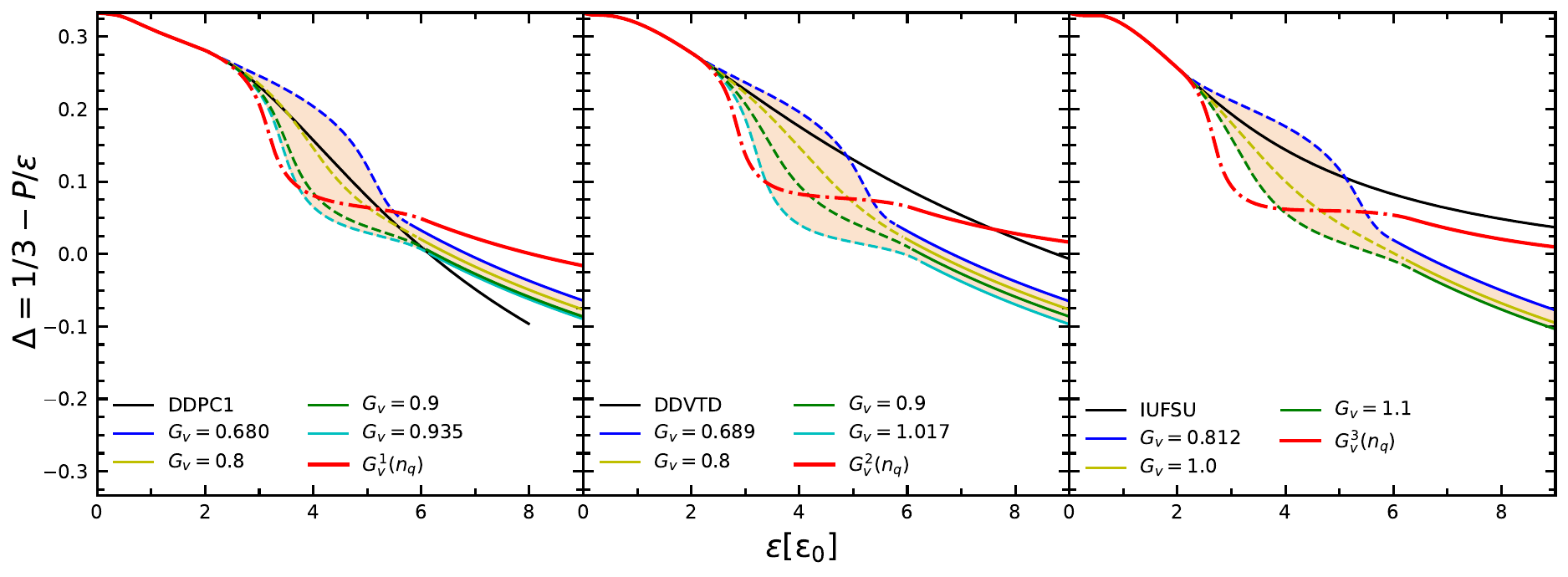}
        \caption{The trace anomaly as a function of energy density with crossover and pure hadronic EOSs.}
        \label{fig7}
    \end{figure}

    The trace anomaly, $\Delta = \frac{1}{3} - \frac{P}{\epsilon}$, is commonly used as a measure of conformality \cite{Fujimoto2022,Zhou2024}. Under the requirements of thermodynamic stability and causality, it must satisfy the condition $-\frac{2}{3} < \frac{1}{3} - \frac{P}{\epsilon} < \frac{1}{3}$. As shown in Fig.~\ref{fig7}, when the crossover is taken into account, the trace anomalies of the hybrid EOSs corresponding to different quark vector coupling strengths exhibit significant differences. A stiffer quark EOS leads to a rapid decrease of the trace anomaly, whereas a softer quark EOS shows the opposite trend. At high densities, the differences among EOSs with different values of $G_v$ become much smaller. For the EOSs generated by the DDNJL models, $\Delta$ decreases rapidly at low densities and tends to zero in the high-density region.
   
    \begin{figure}[htbp]
        \centering
        \includegraphics[width=1\textwidth]{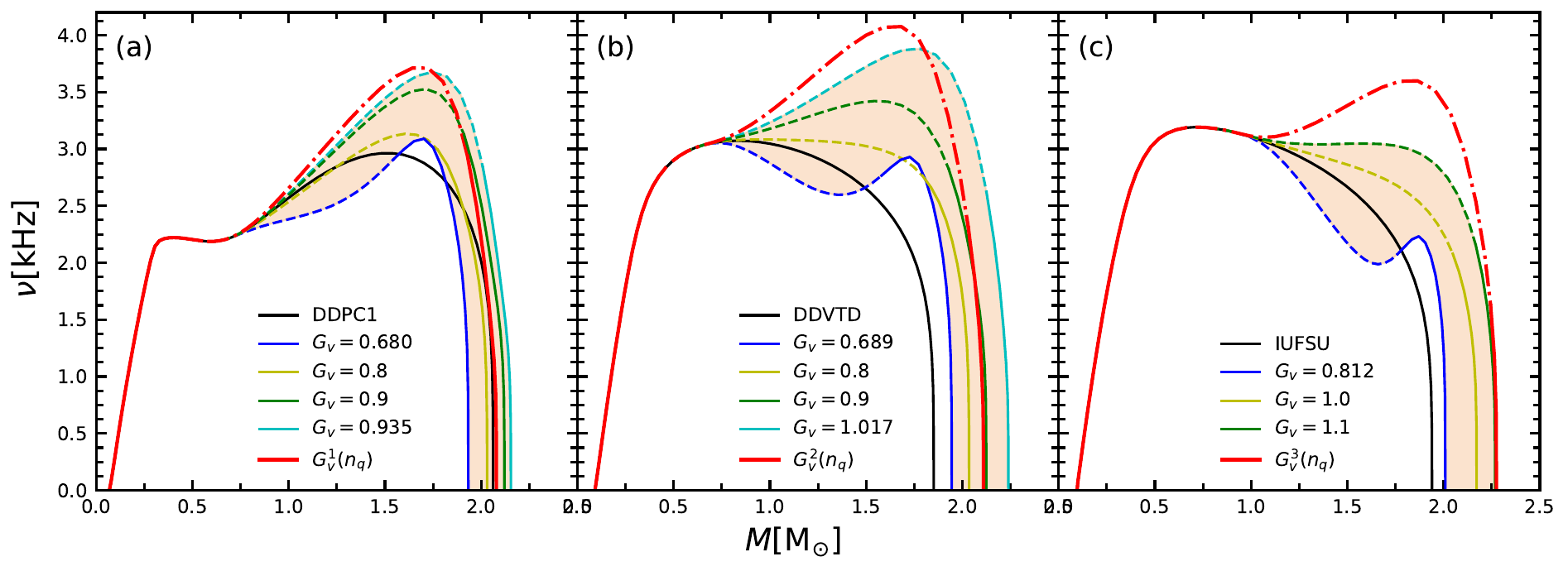}
        \caption{ The radial oscillation frequency as a function of  NS mass at fundamental mode with present EOSs.}
        \label{fig8}
    \end{figure}

    With the advancement of gravitational-wave detection techniques, gravitational waves generated by oscillations of NSs can serve as probes of their internal structure \cite{Passamonti2006,Punturo2010,Chirenti2019,Andersson1996,Andersson1998,Geng2025}. The radial oscillation frequencies, $\nu$, of crossover and pure hadronic EOSs for the fundamental mode, obtained with different parameter sets, are shown in Fig.~\ref{fig8}. For pure hadronic EOSs, except for the DDPC1 set, the radial oscillation frequency first increases with the NS mass, reaches a maximum around $0.5M_\odot$, and then decreases. In contrast, for the crossover EOSs, the oscillation frequency tends to reach its maximum at a larger mass, around $1.8M_\odot$. For crossover EOSs with stronger quark vector coupling, the radial oscillation frequency becomes larger, with an increase of about $20\%$–$40\%$. Moreover, the DDNJL model can generate a higher maximum frequency. As a result, a two-peak structure appears in the $\nu$–$M$ plane. This behaviour may provide a potential observational signature to distinguish pure neutron stars from hybrid stars through measurements of their radial oscillation frequencies.
    
  Finally, the global properties of NSs—including the central pressure at $1.4M_\odot$ ($P_{1.4}$), the radius at $1.4M_\odot$ ($R_{1.4}$), the dimensionless tidal deformability ($\Lambda_{1.4}$), the fundamental radial oscillation frequency at $1.4M_\odot$ ($\nu_{1.4}$), the mass range of the crossover region $M_{\rm cross}$, the maximum mass ($M_{\max}$), and the corresponding radius ($R_{\max}$)—obtained with the three hadronic and crossover EOSs are summarized in Table~\ref{table2}.
  
  For the DDPC1 parameter set, the maximum NS mass is $2.06M_\odot$ for the pure hadronic EOS. When a weak quark vector coupling is included in the crossover scenario, $M_{\max}$ decreases to $1.93M_\odot$. As the vector coupling is increased to $G_v = 0.935G_s$, the maximum mass rises to $2.15M_\odot$. Similar trends are observed for $R_{1.4}$, $\Lambda_{1.4}$, $\nu_{1.4}$, and $R_{\max}$. {As $G_v$ increases, the NS mass range of the crossover region expands from $0.69-1.72M_\odot$ to $0.69-2.11M_\odot$, and the onset of pure quark matter is delayed.} The NS properties obtained with the DDVTD and IUFSU parameter sets exhibit behaviour consistent with that of DDPC1, except for the maximum mass. For the pure hadronic EOSs, $M_{\max}$ in the DDVTD and IUFSU sets takes the smallest values among the three models. As $G_v$ increases, the maximum mass grows significantly, reflecting the fact that these EOSs are relatively soft in the absence of quark matter.
    
    \begin{table*}[htbp]
    	\footnotesize
    	\centering
    	\caption{Properties of NSs obtained by the hadronic and crossover EOSs.}\label{table2}
    	\scalebox{0.8}{
    		\begin{tabular}{r|cccccccc}
    			\hline\hline
    			&$G_v[G_s]$ ~&$P_{1.4}[\mathrm{MeV/fm^3}]$ ~&$R_{1.4}[\mathrm{km}]$ ~&$\Lambda_{1.4}$ ~&$\nu_{1.4}[\mathrm{kHz}]$ ~&$M_{\rm {cross}}[M_{\odot}]$ ~&$M_{max}[M_{\odot}]$ ~&$R_{max}[\mathrm{km}]$ \\
    			\hline 
    			DDPC1 ~& - ~&92.86 ~&11.65 ~&268.96 ~&2.94  ~& - ~&2.06 ~&10.08 \\
    			~&$0.680$ ~&122.97 ~&11.22 ~&193.47  ~&2.68 ~& 0.69-1.72 ~&1.93 ~&9.81 \\
    			~&$0.8$ ~&94.02 ~&11.62 ~&263.16  ~&3.00  ~& 0.69-1.91 ~&2.03 ~&10.33 \\
    			~&$0.9$ ~&84.99 ~&11.76 ~&293.12 ~&3.23  ~& 0.69-2.06 ~&2.12 ~&10.77 \\
    			~&$0.935$ ~&83.13 ~&11.79 ~&300.28 ~&3.28  ~& 0.69-2.11 ~&2.15 ~&10.93 \\
    			~&$G^1_v(n_B)$ ~&78.27 ~&11.88 ~&320.76 ~&3.40  ~& 0.69-2.05 ~&2.07 ~&11.18 \\
    			\hline
    			DDVTD ~& - ~&92.40 ~&11.57 ~&271.71 ~&2.77  ~& - ~&1.85 ~&9.94 \\
    			~&$0.689$  ~&110.00 ~&11.32 ~&223.51 ~&2.60 ~&0.66-1.75 ~&1.94 ~&9.86 \\
    			~&$0.8$ ~&85.91 ~&11.66 ~&292.10 ~&3.05 ~&0.66-1.92 ~&2.03 ~&10.34 \\
    			~&$0.9$ ~&77.81 ~&11.79 ~&321.78 ~&3.38 ~&0.66-2.07 ~&2.12 ~&10.79 \\
    			~&$1.017$ ~&73.64 ~&11.86 ~&338.58 ~&3.62 ~&0.66-2.22 ~&2.24 ~&11.32 \\
    			~&$G^2_v(n_B)$ ~&68.54 ~&11.95 ~&365.82 ~&3.88 ~&0.66-2.10 ~&2.11 ~&11.63 \\
    			\hline
    			IUFSU ~& -  ~&58.35  ~&12.58  ~&483.44  ~&2.78  ~& - ~&1.94 ~&11.23 \\
    			~&$0.812$  ~&63.91 ~&12.48 ~&448.16 ~&2.38 ~&0.94-1.88 ~&2.01 ~&10.35 \\
    			~&$1.0$  ~&57.42 ~&12.60 ~&486.70 ~&2.91 ~&0.94-2.14 ~&2.17 ~&11.18 \\
    			~&$1.1$  ~&56.15 ~&12.62 ~&496.50 ~&3.04 ~&0.94-2.26 ~&2.26 ~&11.63 \\
    			~&$G^3_v(n_B)$  ~&54.06 ~&12.66 ~&510.59 ~&3.28 ~&0.94-2.27 ~&2.27 ~&12.42 \\
    			\hline\hline     
    	\end{tabular}}
    \end{table*}
    
	\section{Summary and perspectives}\label{IV}
    
 In this work, the hadron--quark crossover in neutron star matter was studied by consistently combining RMF hadronic models with the NJL description of quark matter. The vector and diquark coupling constants in the NJL model were constrained using pQCD calculations at high density through a likelihood-based scale-averaging procedure, together with constraints from latest NS mass-radius observations and the causality requirement on the speed of sound.
    
    It was found that the diquark coupling was strongly constrained to $H \simeq 1.5G_s$ by the combined requirements from the sound speed behaviour and NS mass--radius measurements. In contrast, the allowed range of the vector coupling $G_v$ was shown to depend on both the hadronic EOS and the quark sector, with pQCD providing a robust upper bound of $G_v \lesssim 1.1G_s$. For softer hadronic EOSs, the hadron--quark crossover was demonstrated to substantially enhance the maximum NS mass, with increases of up to about $\sim 20\%$, while remaining consistent with current observational constraints. {When the crossover was considered, the fundamental radial oscillation frequency may rise rapidly in the intermediate-mass NSs, or a double-peaked structure may emerge, which could provide a new probe for exploring the internal structure of neutron stars.}
    
    To better satisfy the pQCD constraints, the vector quark coupling strength in the NJL model was extended to be density-dependent, leading to a peak in the speed of sound at relatively low densities. A systematic comparison between constant and density-dependent vector couplings in the NJL model was also performed. {It is demonstrated that the density-dependent vector coupling is capable of generating a shorter-ranged and faster stiffening in the crossover region;    accordingly, the resulting neutron-star radii, tidal deformabilities, and radial oscillation frequencies fall outside the envelope established by the weak and strong constant vector coupling scenarios.} This behaviour highlighted the nontrivial role of the density dependence of quark interactions in determining neutron-star observables. In addition, characteristic features in the sound speed and trace anomaly were identified, including a pronounced peak in $c_s^2$ in the crossover region and a rapid approach toward the conformal limit at high density.
    
    In future studies, the present framework could be extended to include finite-temperature effects, rotation, and nonradial oscillation modes, which would be relevant for applications to binary mergers and post-merger remnants. A more systematic treatment of uncertainties in both the hadronic and quark matter sectors, together with improved pQCD inputs, may further strengthen the robustness of the conclusions. {We expect that future multi-messenger observations will be able to detect the distinctive signatures of quark matter in NS identified in this work, such as the rapid increase in oscillation frequencies for intermediate-mass neutron stars and the potential double-peak structure in radial oscillations.}

	\section{Acknowledgments}
	This work was supported in part by the National Natural Science Foundation of China Nos. 12175109, 12475149, 12275141, and the Guangdong Basic and Applied Basic Research Foundation (Grant  No. 2024A1515010911). 
	
	\section{Data Availability}
	The  python code and data that support the findings of this article are openly available in zenodo \cite{zenodo}.

	\bibliographystyle{apsrev4-2}
	\bibliography{ref}
	
\end{document}